\documentstyle[prl,aps]{revtex}

\begin{document}
\author{Jian-Qi Shen $^{1,}$$^{2}$ \footnote{E-mail address: jqshen@coer.zju.edu.cn} and Shao-Long He $^{2}$}
\address{$^{1}$  Center for Optical
and Electromagnetic Research, Zhejiang University,
Hangzhou 310027, P. R. China\\
$^{2}$Zhejiang Institute of Modern Physics and Department of
Physics, Zhejiang University, Hangzhou 310027, P. R. China}
\date{\today }
\title{Geometric Phases of Electrons Due to the Spin-rotation Coupling \\in Rotating C$_{60}$ Molecules}
\maketitle

\begin{abstract}
The rapidly rotational motion of C$_{60}$ molecules will provide
us with an ingenious way to test Mashhoon's spin-rotation
coupling. Geometric phases arising from the time-dependent
spin-rotation coupling of electrons in the rotating C$_{60}$
molecule is considered in the present Letter. It is shown that
geometric phases of electrons in C$_{60}$ molecules may be
measured through the photoelectron spectroscopy of C$_{60}$. A
physically interesting fact that the information about rotation
and precession of C$_{60}$ molecules in the orientational ordered
(or disordered) phase may be read off from the photoelectron
spectroscopy of C$_{60}$ is also demonstrated.


\end{abstract}
The information on the rotational dynamics of $C_{60}$ molecule in
condensed phases have been obtained from nuclear-magnetic
resonance (NMR) studies\cite{Yannoni}. Both NMR spectroscopy and
quasielastic neutron scattering experiments\cite{Neumann}
indicated the rapid rotation of C$_{60}$ molecules whose
rotational correlation time may be picoseconds in the
orientationally disordered phase. Historically, many researches
were in connection with the molecular-dynamics of $C_{60}$
rotation\cite{Heiney}. Heiney {\it et al.} found from x-ray
diffraction and calorimetrical measurements that solid C$_{60}$
exhibits a phase transition near $249$ K from a simple cubic
structure at low temperatures to a face-centered-cubic structure
at high temperatures\cite{Heiney,Johnson,Kiefl}. These studies
suggest that the phase above the transition temperature is
characterized by free rotation or rotational diffusion and that
the phase below the transition is characterized by jump rotational
diffusion between symmetry-equivalent orientations. The
correlation time for both phases (orientationally ordered phase
below $249$ K and orientationally disordered phase above $249$ K)
were measured to fit with an Arrhenius law, which leads to the
fact that in the low-temperature phase, the correlation time $\tau
$ increases by a factor of $\sim 40$. For example,  $\tau $ varies
from $0.44$ to $0.29$ ns as the temperature varies between $241$
and $254$ K\cite{Johnson}. In order to investigate the dynamics of
C$_{60}$ rotation, Johnson {\it et al.} performed detailed
measurements of the reorientational correlation time for solid
C$_{60}$ over the temperature range $240$ to $331$ K and again
showed that the correlation time satisfies the Arrhenius
behavior\cite{Johnson}. Cheng {\it et al.} presented
molecular-dynamics investigation of orientational freezing in pure
solid C$_{60}$ based on a pairwise-additive atom-atom
intermolecular potential\cite{Heiney}. Kiefl {\it et al.} reported
the study of the molecular dynamics and electronic structure of
$\mu ^{+}$-C$_{60}$ radical in a highly crystalline sample of pure
C$_{60}$ and showed a signal at room temperature which is a $\mu
^{+}$-C$_{60}$ radical in crystalline sample C$_{60}$ undergoing a
quasi-free rotation\cite{Kiefl}. Thus it follows that the
molecular-dynamics of C$_{60}$ rotation is of great importance,
since it is related close to the molecular thermal motion, phase
transition and crystal structure of solid C$_{60}$.

In this Letter we propose a new approach to the investigation of
molecular-dynamics of C$_{60}$. We suggest that both the
spin-rotation coupling of electrons and the consequent geometric
phases can provide us with an insight into the rotational
dynamics, intermolecular interaction and thermal motion of
C$_{60}$ molecules.

Basically speaking, the spin-rotation coupling considered here is
one of the gravitational effects since the nature of the inertial
force ({\it e.g.}, the Coriolis force) is the gravitational force
according to the principle of equivalence. Mashhoon showed that a
particle with an intrinsic spin possesses a gravitomagnetic moment
that can be coupled to the gravitomagnetic fields\cite{Mashhoon0}.
By using the coordinate transformation of gravitomagnetic vector
potentials in Kerr metric from the fixed frame of reference to the
rotating frame\cite{Shen}, it is readily verified that the
rotating frequency, $\vec { \omega}$, of a rotating frame relative
to the inertial frame of reference can be regarded as a
gravitomagnetic field\cite{Shen}. The Hamiltonian of spin-rotation
coupling, $H^{\rm s-r}=\vec { \omega} \cdot\vec { S}$, can be
obtained\cite{Mashhoon0,Shen,Mashhoon1,Mashhoon2} by making use of
the Dirac equation in the curved spacetime\cite{Hehl}, where $\vec
{ S}$ denotes the spin operator of a spin-$\frac{1}{2}$ particle.

Fleming {\it et al.} have considered the electron-spin relaxation
of C$_{60}$ induced by the coupling of the electron spin to the
molecular-rotational-angular momentum\cite{Fleming}. Since in some
certain cases the rotating frequency ({\it e.g.}, $10^{11}$ rad/s
in orientationally disordered phases \cite{Johnson}) of C$_{60}$
molecule can be compared to the electron Larmor frequency in a
magnetic field arising from molecule rotation, it is necessary to
consider the gravitational analogue of Fleming's case. Moreover,
in some cases such as in the low-temperature phase, the precession
of C$_{60}$ molecules around cones is faster than the molecular
rotation (see, for example, in the following evaluation) and
therefore it is also essential to deal with the precessional
motion of C$_{60}$ molecules and the consequent geometric phases
of electrons in molecules.

First we consider the geometric phase of electrons arising from
the time-dependent spin-rotation coupling in the rotating C$_{60}$
molecule which undergoes a precessional motion. Note that the
gravitational Aharonov-Bohm effect (also termed Aharonov-Carmi
effect\cite{Anandan}) due to the Lorentz magnetic force ({\it i.
e.}, the Coriolis force) can be ignored in our following
consideration, since the Berry's phase, $2m_{e}\vec { \omega}
\cdot \vec { A}$ with $\vec { A}$ being the area vector surrounded
by a closed path along which the electron moves, appearing in the
gravitational Aharonov-Bohm effect is negligibly small compared
with the geometric phase due to the interaction of electron spin
with the rotating frame (spin-rotation coupling). In what follows
the precession of angular momentum of rotating C$_{60}$ molecules
is taken into account. Since the interaction between C$_{60}$
molecules in solid C$_{60}$ is the Van der Waals potential, the
coupling is relatively weak and therefore the magnitude of $\vec {
\left| \omega \right|} $ does not easily alter, except that it
gives rise to the variations of the direction of $\vec { \omega
}$. By the aid of the above assumptions ({\it e.g.}, $\frac{\rm
d}{{\rm d}t}\vec { \omega} ^{2}=0$), the phenomenological equation
for the rotational motion of the C$_{60}$ molecule weakly coupled
to its neighbors is derived as $\frac{\rm d}{{\rm d}t}\vec {
L}=\vec { B}\times \vec { L}$ with the effective gravitomagnetic
field strength $\vec { B}$ (which leads to the precessional motion
of C$_{60}$ molecules) being $\frac{\vec { \omega} \times
\dot{\vec { \omega}}}{\omega ^{2}}$, where dot denotes the
derivative of $\vec { \omega}$ with respect time $t$; the angular
momentum $\vec { L}$ of the C$_{60}$ molecule is defined to be
$\vec { L}=I\vec { \omega} $ with the moment of inertia, $I$,
being approximately $\frac{2}{3}ma^{2}$. Here $m$ and $a$ ($0.355$
{\rm $\mathring{A}$}) denote the mass and radius of C$_{60}$
molecule, respectively. Note that here the rotational angular
velocity $\vec { \omega}$ of C$_{60}$ molecules acted upon by
intermolecular interactions is time-dependent due to the molecular
thermal motion and thermal fluctuations in solid C$_{60}$. So, the
Hamiltonian describing the coupling of spin of electrons (such as
those in the delocalized $\pi $ bond which has the conjugation
effect) to the time-dependent $\vec { \omega}(t)$ is given $
H^{\rm s-r}(t)=\vec { \omega} (t)\cdot\vec { S}$ and the
time-dependent Schr\"{o}dinger equation governing this electron
spin-rotation coupling is

\begin{equation}
H^{\rm s-r}(t)\left| \psi _{\sigma }(t)\right\rangle =i\hbar
\frac{\partial }{\partial t}\left| \psi _{\sigma
}(t)\right\rangle.    \label{eq2}
\end{equation}

For convenience, we set $\vec{ \omega} (t)=\omega _{0}\left[ \sin
\theta (t)\cos \varphi (t),\sin \theta (t)\sin \varphi (t),\cos
\theta (t)\right] $ where $\omega _{0}$ is time-independent and
$\theta (t)$ and $\varphi (t)$ stand for the angle displacements
in the spherical polar coordinate system. In order to exactly
solve the time-dependent Schr\"{o}dinger equation (\ref{eq2}), in
what follows use is made of the Lewis-Riesenfeld invariant
theory\cite{Lewis} and the invariant-related unitary
transformation formulation\cite{Gao,Gao2}. So, according to the
Lewis-Riesenfeld theory\cite{Lewis} an invariant $I(t)$ that
satisfies the Liouville-Von Neumann equation $\frac{\partial
}{\partial t}I(t)+\frac{1}{i\hbar }\left[ I(t),H^{\rm
s-r}(t)\right] =0$ should be constructed in terms of the spin
operator $S_{\pm }$ and $S_{3}$ with $S_{\pm }=S_{1}\pm iS_{2}$.
Thus the invariant $I(t)$ is of the form $I(t)=\frac{1}{2}\sin
\lambda (t)\exp \left[ -i\gamma (t)\right] S_{+}+\frac{1}{2}\sin
\lambda (t)\exp \left[ i\gamma (t)\right] S_{-}+\cos \lambda
(t)S_{3}$, where the time-dependent parameters, $\lambda (t)$ and
$\gamma (t)$, are determined by the following two auxiliary
equations

\begin{equation}
\dot{\lambda}=\omega _{0}\sin \theta \sin (\varphi -\gamma), \quad
\dot{\gamma}=\omega _{0}\left[ \cos \theta -\sin \theta \cot
\lambda \cos (\varphi -\gamma)\right] \label{eq3}
\end{equation}
with dot denoting the time derivative of $\lambda (t)$ and $\gamma
(t)$. Note that insertion of expressions for $I(t)$ and $H^{\rm
s-r}(t)$ into the Liouville-Von Neumann equation yields
Eq.(\ref{eq3}). In accordance with the Liouville-Von Neumann
equation, the invariant $I(t)$ possesses {\it time-independent}
eigenvalues, which enables us to obtain the exact solutions of the
time-dependent Schr\"{o}dinger equation (\ref{eq2}), for it is
readily verified that the particular solutions of (\ref{eq2}) is
different from the eigenstates of the invariant $I(t)$ only by a
time-dependent $c$- number factor\cite{Lewis}. If once the
eigenvalue equation of $I(t)$ is exactly solved, then the
solutions of the time-dependent Schr\"{o}dinger equation
(\ref{eq2}) is easily obtained. By utilizing a unitary
transformation operator\cite{Gao,Shen2,Shen3} $V(t)=\exp \left[
\beta S_{+}-\beta ^{\ast }S_{-}\right] $ with $\beta
(t)=-\frac{\lambda (t)}{2}\exp \left[ -i\gamma (t)\right] $ and
$\beta ^{\ast }(t)=-\frac{\lambda (t)}{2}\exp \left[ i\gamma
(t)\right] $, the {\it time-dependent} invariant $I(t)$ can be
transformed into a {\it time-independent} one, {\it i.e.},
$I_{V}\equiv V^{\dagger }(t)I(t)V(t)=S_{3}$, where the
time-independent eigenvalue of $S_{3}$ is $\sigma =\pm
\frac{1}{2}\hbar $ for the electron.

Since $\frac{\partial }{\partial t}I_{V}=0$, the Liouville-Von
Neumann equation under this unitary transformation may be
rewritten $\left[ I_{V},H_{V}^{\rm s-r}(t)\right]=0 $. It follows
that $H_{V}^{\rm s-r}(t)$ depends only on the third component
$S_{3}$ of the spin operator of the electron. Further calculation
yields

\begin{equation}
H_{V}^{\rm s-r}(t)\equiv V^{\dagger }(t)[ H^{\rm s-r}(t)-i\hbar
\frac{\partial }{\partial t}] V(t)=\omega _{0}\left[ \cos \lambda
\cos \theta +\sin \lambda \sin \theta \cos (\gamma -\varphi
)\right] S_{3}+\dot{\gamma}\left(1-\cos \lambda \right) S_{3}.
 \label{eq4}
\end{equation}
This, therefore, means that the particular solution $\left| \psi
_{\sigma }(t)\right\rangle _{V}$ of the time-dependent
Schr\"{o}dinger equation $H_{V}^{\rm s-r}(t)\left| \psi _{\sigma
}(t)\right\rangle _{V}=i\hbar \frac{\partial }{\partial t}\left|
\psi _{\sigma }(t)\right\rangle _{V}$ with $\left| \psi _{\sigma
}(t)\right\rangle _{V}=V^{\dagger }(t)\left| \psi _{\sigma
}(t)\right\rangle $ is different from the eigenstate of $I_{V}$
only by a time-dependent $c$- number factor $\exp \left\{
\frac{1}{i\hbar }[ \phi _{\sigma }^{({\rm d})}(t)+\phi _{\sigma
}^{({\rm g})}(t)]\right\} $, where the nonadiabatic noncyclic
dynamical phase is
\begin{equation}
\phi _{\sigma }^{({\rm d})}(t)=\sigma \int_{0}^{t}\omega
_{0}\left\{ \cos \lambda (t^{\prime })\cos \theta (t^{\prime
})+\sin \lambda (t^{\prime })\sin \theta (t^{\prime })\cos \left[
\gamma (t^{\prime })-\varphi (t^{\prime })\right] \right\} {\rm
d}t^{\prime }  \label{eq5}
\end{equation}
and the nonadiabatic noncyclic geometric phase is
\begin{equation}
\phi _{\sigma }^{({\rm g})}(t)=\sigma \int_{0}^{t}\left\{
\dot{\gamma}(t^{\prime })\left[ 1-\cos \lambda (t^{\prime
})\right] \right\} {\rm d}t^{\prime }. \label{eq6}
\end{equation}
If the adiabatic cyclic evolution process is taken into
consideration, then it is verified that the present calculation is
self-consistent. This may be illustrated as follows: in the
adiabatic cyclic case, where the parameters $\lambda =\theta={\rm
const.} $, $\gamma =\varphi $, and the precessional frequency
$\dot{\gamma}=\Omega $ is very small, the nonadiabatic geometric
phase (\ref{eq6}) in one cycle ($T=\frac{2\pi }{\Omega }$) is
reduced to the expression for Berry's adiabatic cyclic topological
phase $\phi _{\sigma }^{({\rm g})}(T)=2\pi \sigma\left(1-\cos
\theta \right) $\cite{Berry}, where $2\pi\left(1-\cos \theta
\right) $ is a solid angle over the parameter space of the
C$_{60}$ rotational frequency $\vec{\omega}(t)$. This fact implies
that geometric phases possess the topological and global
properties of time evolution in time-dependent quantum
systems\cite{Shen2,Berry}.

If the eigenstate of $I_{V}$ corresponding to the eigenvalue
$\sigma $ is $\left| \sigma \right\rangle $, then the eigenstate
of the invariant $I(t)$ is $V(t)\left| \sigma \right\rangle $.
Hence, in accordance with the Lewis-Riesenfeld invariant theory,
the particular solution corresponding to $\sigma $ of
Eq.(\ref{eq2}) reads

\begin{equation}
\left| \psi _{\sigma }(t)\right\rangle =\exp \left\{
\frac{1}{i\hbar }\left[ \phi _{\sigma }^{({\rm d})}(t)+\phi
_{\sigma }^{({\rm g})}(t)\right] \right\} V(t)\left| \sigma
\right\rangle.  \label{eq7}
\end{equation}

The influence of spin-rotation coupling and geometric phases
discussed above on the photoelectron spectroscopy in the C$_{60}$
molecule deserves considerations. The total Hamiltonian of an
electron in the C$_{60}$ molecule acted upon by the external
perturbation ({\it e.g.}, the radiation fields) can be written as
$H=H_{0}+H^{\rm s-r}(t)+H^{\prime }(t)$, where $H_{0}$ represents
the Hamiltonian of the electron in the C$_{60}$ molecule when no
spin-rotation coupling and external perturbation exist, and
$H^{\prime }(t)$ describes the external perturbation acting on the
electrons. Since $H_{0}$ is assumed to be time-independent, the
eigenvalue equation of $H_{0}$ may be written $H_{0}\exp \left[
\frac{1}{i\hbar }\epsilon _{n,\sigma }t\right] \left| \phi
_{n,\sigma }\right\rangle =\epsilon _{n,\sigma }\exp \left[
\frac{1}{i\hbar }\epsilon _{n,\sigma }t\right] \left| \phi
_{n,\sigma }\right\rangle$. The initial state $\left| \psi
_{n,\sigma }(t=0)\right\rangle $ of Eq.(\ref{eq2}) may be taken to
be $\left| \phi _{n,\sigma }\right\rangle $. Thus, the particular
solution of the time-dependent equation $ \left[ H_{0}+H^{\rm
s-r}(t)\right] \left| \Phi _{n,\sigma }(t)\right\rangle =i\hbar
\frac{\partial }{\partial t}\left| \Phi _{n,\sigma
}(t)\right\rangle $ is given $ \left| \Phi _{n,\sigma
}(t)\right\rangle =\exp \left\{ \frac{1}{i\hbar }\left[ \phi
_{\sigma }(t)+\epsilon _{n,\sigma }t\right] \right\} V(t)\left|
\phi _{n,\sigma }\right\rangle $ with $\phi _{\sigma }(t)=\phi
_{\sigma }^{({\rm d})}(t)+\phi _{\sigma }^{({\rm g})}(t)$. Note,
however, that here the magnetic interactions such as spin-orbit
and spin-spin couplings in $H_{0}$ has been combined into $H^{\rm
s-r}(t)$ ({\it i.e.}, this leads to the commutation relation
$\left[ H_{0},H^{\rm s-r}(t)\right] =0$) in order that the exact
particular solution can be obtained conveniently via the above
unitary transformation method. The solution of the time-dependent
Schr\"{o}dinger equation $\left[ H_{0}+H^{\rm s-r}(t)+H^{\prime
}(t)\right] \left| \Psi (t)\right\rangle =i\hbar \frac{\partial
}{\partial t}\left| \Psi (t)\right\rangle$ associated with the
total Hamiltonian is assumed to be $\left| \Psi (t)\right\rangle
=\sum_{n,\sigma }a_{n,\sigma }(t)\left| \Phi _{n,\sigma
}(t)\right\rangle $. Substitution of this expression into this
time-dependent Schr\"{o}dinger equation yields $i\hbar \frac{{\rm
d}}{{\rm d}t}a_{m,\sigma ^{\prime }}(t)=\sum_{n,\sigma
}a_{n,\sigma }(t)\left\langle \Phi _{m,\sigma ^{\prime
}}(t)\right| H^{\prime }(t)\left| \Phi _{n,\sigma
}(t)\right\rangle$. Further calculation shows that
\begin{equation}
\left\langle \Phi _{m,\sigma ^{\prime }}(t)\right| H^{\prime
}(t)\left| \Phi _{n,\sigma }(t)\right\rangle =\exp [
\frac{1}{i\hbar }\phi _{\rm tot}(m\sigma ^{\prime },n\sigma ;t)]
H_{m\sigma ^{\prime },n\sigma }^{\prime }(t), \label{eq11}
\end{equation}
where $H_{m\sigma ^{\prime },n\sigma }^{\prime }(t)=\left\langle
\phi _{m,\sigma ^{\prime }}\right| V^{\dagger }(t)H^{\prime
}(t)V(t)\left| \phi _{n,\sigma }\right\rangle $ and $\phi _{\rm
tot}(m\sigma ^{\prime },n\sigma ;t)=\left[ \phi _{\sigma
}(t)+\epsilon _{n,\sigma }t\right] -\left[ \phi _{\sigma ^{\prime
}}(t)+\epsilon _{m,\sigma ^{\prime }}t\right] $.

If the initial state is $\left| \Phi _{k,\sigma }\right\rangle $,
{\it i.e.}, $a_{k,\sigma }(t=0)=1$, then $a_{m,\sigma ^{\prime
}}(t)=\frac{1}{i\hbar }\int_{0}^{t}H_{m\sigma ^{\prime },k\sigma
}^{\prime }(t^{\prime })\exp \left[ \frac{1}{i\hbar}\phi _{\rm
tot}(m\sigma ^{\prime },k\sigma ;t^{\prime })\right] {\rm
d}t^{\prime }$, and the transition probability from the state
$\left| \Phi _{k,\sigma }\right\rangle $ to $\left| \Phi
_{m,\sigma ^{\prime }}(t)\right\rangle $ is $W_{k\sigma
\rightarrow m\sigma ^{\prime }}=a_{m,\sigma ^{\prime }}^{\ast
}(t)a_{m,\sigma ^{\prime }}(t)$.

Note that the exact solutions of the auxiliary equations
(\ref{eq3}) are often of complicated form.  As an illustrative
example, here we consider a physically interesting solution to
Eq.(\ref{eq3}), which stands for a typical case where the
precessional frequency $\dot{\varphi}$ is constant (denoted by
$\Omega $) and the nutational frequency $\dot{\theta}$ vanishing.
The explicit expression for this simple solution is $\gamma (t)
=\varphi (t)=\Omega t,\quad \dot{\lambda}=\dot{\theta}=0$, where
the time-independent $\Omega =\frac{\omega _{0}\sin \left(\lambda
-\theta \right) }{\sin \lambda }$. Here the C$_{60}$ molecule
precesses at an angular velocity $\Omega $ about the $z$-axis
($\vec { \omega}$ deviates from $z$-axis by a constant angle
$\theta$). Thus it follows that in this case the nonadiabatic
noncyclic dynamical phase and geometric phase of the electron are
proportional to time $t$, {\it i.e.}, $\phi _{\sigma }^{({\rm
d})}(t)=\left[ \omega _{0}\sigma \cos \left(\lambda -\theta
\right) \right] t$ and $\phi _{\sigma }^{({\rm g})}(t)=\Omega
\sigma \left(1-\cos \lambda \right) t$, respectively, and the
total phase difference $\phi _{\rm tot}(m\sigma ^{\prime },k\sigma
;t)$ in transition matrix element $\left\langle \Phi _{m,\sigma
^{\prime }}(t)\right| H^{\prime }(t)\left| \Phi _{k,\sigma
}(t)\right\rangle $ is therefore of the form

\begin{equation}
\phi _{\rm tot}(m\sigma ^{\prime },k\sigma ;t)=\left\{{ \left(
{\sigma -\sigma ^{\prime }}\right) \left[ \omega _{0}\cos
\left(\lambda -\theta \right) +\Omega \left(1-\cos \lambda \right)
\right] +\left( \epsilon _{k,\sigma }-\epsilon _{m,\sigma ^{\prime
}}\right) }\right\} t.          \label{eq12}
\end{equation}

It follows from the expression (\ref{eq12}) that the photon energy
absorbed or emitted in the transition process from $\left| \Phi
_{k,\sigma }(t)\right\rangle $ to $\left| \Phi _{m,\sigma ^{\prime
}}(t)\right\rangle $ is shifted by $\left(\sigma -\sigma ^{\prime
}\right) \left[ \omega _{0}\cos \left(\lambda -\theta \right)
+\Omega \left(1-\cos \lambda \right) \right] $ due to the electron
spin-rotation coupling and the consequent geometric-phase effect.
Apparently, the energy shifted by the dynamical phase and
geometric phase implies the information ({\it i.e.}, $\omega
_{0}$, $\Omega$ and $\theta $) about the rotational motion and
precession of the C$_{60}$ molecule in condensed phases. This,
therefore, means that it is possible for the information on the
molecular rotation and precession to be read off in the
photoelectron spectroscopy.

Let us evaluate the precessional frequency $\Omega $ and the
effective gravitomagnetic field $\vec{B}=\frac{\vec {\omega}
\times \dot {\vec {\omega}}}{\omega ^{2}}$. At $283$ K it is
measured that the molecular reorientational correlation time,
$\tau $, is $9.1$ picoseconds that is three times as long as the
calculated correlation time $\tau $ ($\equiv \frac{3}{5}\left(
\frac{I}{k_{B}T}\right) ^{\frac{1}{2}}$ with $k_{B}$ and $T$ being
the Boltzmann's constant and the absolute temperature,
respectively) for free rotation ({\it i.e.}, the unhindered
gas-phase rotation) at this temperature\cite{Johnson}. The
interaction energy of C$_{60}$ molecule with the angular momentum
$\vec {L}$ acted upon by the effective gravitomagnetic field
$\frac{\vec{\omega} \times \dot {\vec {\omega}}}{\omega ^{2}}$ is
$E=-\frac{\vec {\omega} \times \dot {\vec{\omega}}}{\omega ^{2}}
\cdot \vec {L}$. As is assumed above, the angular velocity of
C$_{60}$ molecule is $\vec {\omega} (t)=\omega _{0}\left(\sin
\theta \cos \Omega t,\sin \theta \sin \Omega t,\cos \theta \right)
$, and the effective gravitomagnetic field is therefore $\vec
{B}=\Omega \sin \theta \left(-\cos \theta \cos \Omega t,-\cos
\theta \sin \Omega t,\sin \theta \right) $ that is apparently
perpendicular to the angular momentum $\vec {L}$. Thus the
magnitude of intermolecular torque acting on the rotating C$_{60}$
molecule is $\left| \vec {M}\right| =\left| \vec {B}\times \vec
{L}\right| =\omega _{0}\Omega I\sin \theta $ with $I\simeq
1.0\times 10^{-43}$ Kg$\cdot $m$^{2}$ \cite{Johnson} being the
moment of inertia of C$_{60}$ molecule. The order of magnitude of
$\left| \vec {M}\right| $ may be approximately equal to (or less
than) the Van der Waals potential energy ($0.001\sim 0.1 $ eV).
Since it follows that in the high-temperature phase
(orientationally disordered phase), $\omega _{0}$ may be $10^{11}$
rad/s, the precessional frequency $\Omega $ is therefore compared
to $\omega _{0}$, {\it i.e.}, $\Omega \simeq \frac{\left| \vec
{M}\right| }{I\omega _{0}} $ ranges from $10^{10}$ to $10^{12}$
rad/s. However, in the low-temperature phase (orientationally
ordered phase), $\omega _{0}$ decreases ({\it e.g.}, $\omega _{0}
\sim 10^{9}$ rad/s) and in turn the precessional frequency $\Omega
$ increases by a factor of $\sim 100$. Thus, the precessional
frequency of C$_{60}$ molecule is much greater than the rotating
frequency $\omega _{0}$ (which means in this case the dynamical
phases due to spin-rotation coupling can be ignored) and the
frequency shift in the transition matrix element due to geometric
phases can be compared to the typical energy of an electron in
solid. It follows that the effects resulting from the
spin-rotation coupling and geometric phases in C$_{60}$ molecules
deserve further investigations both theoretically and
experimentally.

It is apparently seen that investigation of the spin-rotation
geometric phases of electrons in C$_{60}$ molecules is of physical
interest:

(i) it enables us to study the relation between the photoelectron
spectroscopy and the rotational motion of C$_{60}$ molecules.
Moreover, since geometric phases of electrons in one C$_{60}$
molecule depend on other C$_{60}$ molecules (via the effective
gravitomagnetic field strength $\frac{\vec { \omega} \times
\dot{\vec { \omega}}}{\omega ^{2}}$ and hence the intermolecular
torque $\frac{\vec { \omega} \times \dot{\vec { \omega}}}{\omega
^{2}} \times \vec{L}$) and thus imply the information about the
thermal motion and rotational dynamics of C$_{60}$, it is helpful
to analyze the condensed phases (and hence the phase-transition
behavior) of solid C$_{60}$\cite{Laf}.

(ii) for the present, Mashhoon's spin-rotation coupling can be
tested only in microwave experiments\cite{Mashhoon2}, since this
coupling is relatively weak due to the smallness of the rotational
frequency of various rotating frames on the Earth. Fortunately,
here the rotational motion of C$_{60}$ molecule can provide us
with an ingenious way to test this weakly gravitational
(gravitomagnetic) effect. Since the rotational angular velocity,
$\omega _{0}$, of rotating C$_{60}$ molecules is much greater than
that of any rotating bodies on the Earth, the C$_{60}$ molecule is
an ideal noninertial frame of reference for the electrons in the
C$_{60}$ molecule, where the effects resulting from the electron
spin-rotation coupling may be easily observed experimentally.

(iii) as is claimed previously, in addition to the Aharonov-Carmi
geometric phase due to the Coriolis force (gravitomagnetic Lorentz
force)\cite{Anandan}, there exists another geometric phase
associated with gravitational fields arising from the interaction
between the spinning particle and the time-dependent
gravitomagnetic fields. In the present Letter, the time-dependent
gravitomagnetic field strength is just the rotating frequency
$\vec {\omega} (t)$ of C$_{60}$ molecules. It may be reasonably
believed that, from the point of view of equivalence principle in
General Relativity, this geometric phase itself appears to possess
rich physical significance and should therefore be considered in
more detail.

To summarize, we study the C$_{60}$ molecule precession and
geometric phases of electrons due to the time-dependent
spin-rotation coupling in C$_{60}$ molecules. Since in the
orientationally ordered phase the precessional frequency of
C$_{60}$ is fairly great, the effects of geometric phases will be
apparent and even may therefore be read off from the photoelectron
spectroscopy, which enables physicists to investigate the
rotational dynamics and phase transition of C$_{60}$.
Additionally, the present work will make possible a test of
Mashhoon's spin-rotation coupling by measuring the spin-rotation
geometric phases of electrons in C$_{60}$ molecules with rapid
rotation. We hope all these physical phenomena and effects would
be investigated experimentally in the near future.

Acknowledgments This project is supported by the National Natural
Science Foundation of China under the project No. $10074053$ and
the Zhejiang Provincial Natural Science Foundation under the
project No. $100019$.

\end{document}